\documentclass[12pt]{article}
\usepackage{varwidth}
\usepackage{cite}
\usepackage{url}
\usepackage{empheq}
\usepackage{amsthm,amsmath,latexsym,amssymb,amsfonts,amsbsy}
\usepackage{graphicx,lscape,fancyhdr,xcolor,array,stmaryrd,euscript,wrapfig}
\usepackage{hyperref}
\usepackage{tikz}
\usetikzlibrary{decorations.pathmorphing}
\definecolor{darkmagenta}{rgb}{0.55, 0.0, 0.55}
\definecolor{darkblue}{rgb}{0.0, 0.0, 0.55}
\definecolor{darkred}{rgb}{0.7, 0.0, 0.3}
\hypersetup{
	unicode,	
	colorlinks,
	citecolor=darkred,
	linkcolor=darkblue, 
	urlcolor=darkred,
	bookmarksopen=true,
	bookmarksnumbered
	}

 \csname
@addtoreset\endcsname{equation}{section}
\begin{document}
\begin{titlepage}
\setcounter{page}{1}
\begin{center}
\hfill
\vskip 1cm
{\LARGE \bf De Sitter Space and Entanglement}
\vskip 40pt

{\sc Cesar Arias}\footnote{ \href{mailto:a8minutosdelsol@gmail.com}{\texttt{a8minutosdelsol@gmail.com}}},\, 
{\sc Felipe Diaz}\footnote{ \href{mailto:f.diazmartinez@uandresbello.edu}{\texttt{f.diazmartinez@uandresbello.edu}}}
~\&~
{\sc Per Sundell}\footnote{ \href{mailto:per.anders.sundell@gmail.com}{\texttt{per.anders.sundell@gmail.com}}}
\vskip 20pt
{\em Departamento de Ciencias F\'isicas,  
Universidad Andres Bello\\ 
Sazi\'e 2212, Piso 7, Santiago de Chile}
\vskip 40pt 
{\bf Abstract}\\
\end{center}
We argue that the notion of entanglement in de Sitter space arises naturally from the non-trivial Lorentzian geometry of the spacetime manifold, which consists of two disconnected boundaries and a causally disconnected interior.   
In four bulk dimensions, we propose an holographic description of an inertial observer in terms of a thermofield double state in the tensor product of the two boundaries Hilbert spaces, whereby the Gibbons--Hawking formula arises as the holographic entanglement entropy between the past and future conformal infinities. 
When considering the bulk entanglement between the two causally disconnected Rindler wedges, we show that the corresponding entanglement entropy is given by one quarter of the area of the pair of codimension two minimal surfaces that define the set of fixed points of the~dS$_4/\mathbb Z_q$ orbifold.

\end{titlepage}
{\small\tableofcontents }       
\vspace{1 cm}
\section{Introduction}\label{Sec:Intro}
\subsection{The problem of de Sitter entropy}
Astrophysical observations of distant supernov\ae~\cite{1538-3881-116-3-1009, Perlmutter_1999} indicate that our universe is increasingly expanding, driven by a small positive cosmological constant. 
As a consequence of the accelerated expansion, an inertial observer in de Sitter (dS) space is causally connected with only a subregion of the full spacetime, being surrounded by a cosmological horizon. 

As pointed out by Gibbons and Hawking~\cite{Gibbons:1977mu}, dS space exhibits a number of thermodynamic properties that resemble those of a black hole. 
Remarkably, just as in the black hole case \cite{Bekenstein:1973ur, Bekenstein:1974ax, Hawking:1974sw, Hawking:1976de}, inertial observers in dS space detect thermal radiation at a temperature proportional to the inverse of the dS radius, with a corresponding thermodynamic entropy given by one quarter of the area of the cosmological horizon.

An universal description at the microscopic level of the thermal features of dS space remains unknown, mainly because of the lack of a UV-complete theory of gravity on a dS background~\cite{Witten:2001kn, Banks:2000fe, Banks:2001yp}.
However, several proposals have been made. These essentially follow two related approaches, both relying on asymptotic symmetries arguments and their possible central extensions, but differing on the spacetime region where these symmetries are centrally enhanced. 

\paragraph{Near-horizon symmetries.}
One of these approaches has been motivated by Carlip's derivation~\cite{PhysRevD.51.632} of the Bekenstein--Hawking formula for the BTZ black hole~\cite{Banados:1992gq, Banados:1992wn}, and it is based on the algebraic nature and affine (typically Virasoro) extensions of the \emph{near-horizon symmetries}.
Following this approach and making use of the Chern--Simons formulation of three-dimensional gravity, Maldacena and Strominger~\cite{Maldacena:1998ih} showed that the underlying symmetries at the dS horizon corresponds to an $SL(2,\mathbb C)$ current algebra at the boundary of a spatial disk, where the Chern--Simon theory reduces to an $SL(2,\mathbb C)$ Wess--Zumino--Witten model, and where the dS entropy arises as the entropy of a highly excited thermal state, at an energy level roughly given by the (imaginary part of the complex) Chern--Simons level.

Later on, the analysis of the near-horizon symmetries was extended
to any dimension and to arbitrary type of Killing horizon, including the dS horizon. 
By using covariant phase space methods~\cite{PhysRevD.48.R3427, PhysRevD.50.846} and under a suitable set of boundary conditions, Carlip showed~\cite{Carlip:1999cy} that modelling the boundary of a manifold locally as a Killing horizon, the constraint algebra of general relativity acquires a non-trivial central extension given in terms of the gravity coupling constants, which determines the density of states at the boundary and gives rise, via Cardy's formula~\cite{Cardy:1986ie, Bloete:1986qm}, to the quarter of the area formula.

\paragraph{Asymptotic symmetries at spacelike infinity.}
A second route to the dS entropy problem, originally proposed by Strominger in the context of the microstates counting of the three-dimnsional black hole~\cite{Strominger:1997eq} and inspired by the Brown--Henneaux~\cite{Brown:1986nw} construction of aymptotic symmetries of three-dimensional anti de Sitter (AdS) space, treats as the symmetry enhancement  region the \emph{spacelike infinity} of dS space. 
According to the dS/CFT correspondence~\cite{Witten:2001kn, Strominger:2001pn}, its precursors~\cite{Park:1998qk, Hull:1998aa, Banados:1998tb, Park:1998yw, Bousso:1999xy, Bousso:1999cb, Bousso:2000nf, Balasubramanian:2001aa} and refinements~\cite{Klemm:2001ea, Cacciatori:2001un, Balasubramanian:2002aa, Bousso:2002aa, Spradlin:2001nb, Balasubramanian:2002zh, Maldacena:2002vr, Anninos:2011ui}, the microscopic degrees of freedom giving rise to the dS entropy are encoded within a dual Euclidean conformal field theory located at the boundary of global dS space.
Within this framework, the Gibbons--Hawking formula has been correctly reproduced in three dimensions, in which case and under appropriate boundary conditions, the diffeomorphism transformation of the (spacelike) boundary Brown--York stress-energy tensor \cite{PhysRevD.47.1407} produces, via an anomalous Schwarzian derivative, a central charge that is formally equal to the Brown--Henneaux central charge. 

\paragraph{De Sitter entropy as entanglement entropy.}
More recently, following the rationale behind the study of CFT entanglement entropies through holographic techniques~\cite{Ryu:2006bv, Hubeny:2007xt, Casini:2011kv, Dong:2016fnf} in the framework of the AdS/CFT correspondence~\cite{Maldacena:1997re,Gubser:1998bc,Witten:1998qj}, an alternative path driven by the interplay between geometry and entanglement has been taken.
It explores the possibility of understanding the entropy of dS space in terms of the entanglement entropy between disconnected regions of spacetime, each region arguably encoding some underlying quantum field theory states.

Pointing in this direction and making use of the dS/dS correspondence~\cite{Alishahiha:2004md}, which states that gravity on dS space (near the horizon of a certain causal region) is dual to the low energy limit of a conformal field theory coupled to gravity on dS space of one lower dimension, it has been argued~\cite{Dong:2018cuv} that strong interactions between two coupled sectors of the dual theory yield a maximally mixed reduced density matrix. 
This translates into a R\'enyi entropy whose entanglement entropy limit can then be explicitly matched, under certain assumptions, to the Gibbons-Hawking entropy and more precisely so in the case of three-dimensional gravity.

Related investigations link the Gibbons--Hawking entropy to the area of analytically continued extremal surfaces in Euclidean AdS space~\cite{Sato:2015tta}, and to the area of codimension two extremal surfaces stretching between the past and future infinities~\cite{Narayan:2017xca}.
Interestingly, in the latter case, it was further suggested that four-dimensional dS space may be dual to a entangled state (of the thermofield double type) comprising two  copies of the dual conformal field theory, with the dS entropy emerging from the entanglement between the two conformal boundaries~(for related discussions see also \cite{Jatkar:2018aa}). As we shall see in this paper, this proposal seems to be correct.

\subsection{Summary and plan of the paper}
In this work we shall argue that the entropy of dS space has its roots in the connectedness properties of the spacetime, and it should therefore be considered as a consequence of the non-trivial topology of it. 
We consider the entanglement between the two conformal boundaries as well as the entanglement between the two Rindler wedges of the dS interior.
When analyzing the entanglement between the two disconnected boundaries, we find that the Gibbons--Hawking formula arises as the holographic entanglement entropy between the past and future infinities. 
When studying the entanglement between the two Rindler wedges, our findings indicate that the entanglement entropy obeys an area law in terms of the pair of codimension two minimal surfaces that correspond to the set of fixed points of a ${\rm dS}_4/\mathbb Z_q$ orbifold.

The structure and the underlying reasoning of the paper goes as follows. We begin in Section~\ref{Sec:2} by reviewing the basics notions of classical dS geometry that are of importance for the next sections.

In Section~\ref{Sec:3} we focus on the case of four dimensions and introduce a maximally extended set of static coordinates. The advantage of these coordinates is that they cover both, northern and southern Rindler wedges of the dS interior. Accordingly, the extended coordinate system describe the worldline of two antipodal and causally disconnected inertial observers located at the north and south poles of global dS space (at opposites edges of the Penrose diagram; see Figure~\hyperlink{Fig:3}3).

In these coordinates, the dS line element becomes a fibration over a 2-sphere with the warped product form 
$g = S^2 \times_w {\rm dS}^\pm_2$,
where $w$ is a warp factor that depends on the polar coordinate 
of the $S^2$, and ${\rm dS}^\pm_2$ denotes the radially extended 
two-dimensional dS space. 
The $S^2$ factor defines the cosmological horizon and can be consistently
deformed by azimutal identifications. The foliation $g$ thus admits a one 
parameter family of deformations obtained through the orbifold ${\rm dS}_4/\mathbb Z_q$, whose fixed points give rise to a pair of antipodal, codimension-2 defects 
with a local ${\rm dS}^\pm_2$ geometry and endowed with a tension proportional to $(1-q^{-1})$.  
The tensionless limit $q\to1$ corresponds to the undeformed configuration.
The two antipodal defects contain the worldline of the corresponding antipodal 
observer, and hence can be interpreted as the response of the background
geometry to the presence of massive, non-probe observers~\cite{Arias:2019zug} (with a probe limit equivalent to the tensionless limit).

In Section~\ref{Sec:4} we turn to the explicit calculation of entanglement entropy, 
first between the past and future conformal boundaries and then between the 
two interior Rindler wedges.
In the case of boundary entanglement, our derivation relies on the following 
assumptions and symmetry arguments:
\begin{itemize}
\item[$\diamond$] We assume the existence of an holographic duality 
whereby quantum gravity on dS space is dual to two copies of a certain 
conformal field theory, one copy per boundary, and such that a bulk observer 
can be described in terms of a thermofield double state in the tensor product of the 
field theory Hilbert spaces.  As a result, the observer density matrix is thermal.
\item[$\diamond$] Since the two boundaries of dS space have both the topology 
of a 3-sphere, there exists an obvious boundary replica symmetry (given by 
discrete azimutal identifications) that makes the construction of a branched cover
boundary manifold trivial, permitting the usage of the replica method~\cite{0305-4608-5-5-017,Calabrese:2004eu} to 
compute the boundary entanglement entropy.  This replica symmetry extends into 
the bulk as the orbifold ${\rm dS}_4/\mathbb Z_q$.
\item[$\diamond$]  There exists a boundary antipodal map~\cite{Strominger:2001pn, Witten:2001kn} that sends every point on the past 3-sphere to an antipodal point on the future 3-sphere which amounts to formulate the boundary field theory partition function on a single 3-sphere. In the low energy limit, the 3-sphere partition function can be translated to the Euclidean on-shell Einstein gravity action on a single Rindler
wedge (whose topology is that of a 4-sphere in the Euclidean geometry).
\item[$\diamond$] Since the boundary 3-sphere is a space-like boundary, it is not possible to define the temporal evolution of a Cauchy slice containing initial data. However, it is possible to define a codimension one Cauchy-like surface (given by a 2-sphere that result from gluing together two disks along their boundaries) in such a way that the state of the dual theory on the full boundary can be reconstructed \emph{via modular evolution instead of time evolution}. 
\end{itemize}
Tying the above four elements together (and following the standard procedure to
compute $q$-th power of the reduced density matrix) the Gibbons--Hawking formula
arises in the tensionless limit of the $q$-th R\'enyi entropy computed from the bipartition of the Cauchy-like slice. 

\medskip
As for the entanglement between disconnected interior regions,
the analysis does not requires the notion of holography. In this case, assuming that
the two interior Rindler wedges are entangled, we compute the corresponding
entanglement entropy based on the following observations:
\begin{itemize}
\item[$\diamond$] The Euclidean bulk geometry is given by the fibration $g=S^2\times_w S^2$. Hence, there exists a manifest bulk replica symmetry that once implemented deforms 
the geometry of the left $S^2$ factor into that of the $S^2/\mathbb Z_q$
orbifold.  The bulk replica symmetry can thus be thought of as the observers 
back-reaction whose set of fixed points defines a pair of minimal surfaces, 
each with the topology of a 2-sphere. 
\item[$\diamond$]  There exists a bulk antipodal map that sends every point of
the northern Euclidean Rindler wedge (which has the topology of a 4-sphere) to an antipodal point at the southern Euclidean Rindler wedge. This bulk antipodal map
permits to write the full quantum gravity partition with support on a single 4-sphere.
\item[$\diamond$] It is possible to define a Cauchy-like surface (given by a 3-sphere that result from gluing together two 3-balls along their boundaries) whose \emph{modular evolution} reconstruct the state of the bulk theory on the full 4-sphere.
\end{itemize}
Comprising these arguments and using standard semiclassical
techniques to write the quantum gravity partition function in terms of the on-shell
Einstein gravity action, we obtain the entanglement entropy
\begin{equation}
\mathcal S_{\rm E} = \frac{{\rm Area}(\mathcal F)}{4G_4}~,
\end{equation}
where $\mathcal F$ is the set of fixed points of the bulk $\mathbb Z_q$ action.

\medskip
The paper ends in Section~\ref{Sec:5} with a discussion of our results. We also include some potentially interesting directions for future work. 
\section{De Sitter space}\label{Sec:2}
In this section, we shall briefly introduce the key concepts regarding the geometry and thermodynamics of dS space which will be extensively used in the rest of the paper. For comprehensive reviews on the subject see~\cite{Spradlin:aa, HartmanLectures} and references therein. 

$d$-dimensional de Sitter space, denoted by ${\rm dS}_d$, 
is the maximally symmetric Einstein manifold of constant 
positive curvature. ${\rm dS}_d$ can be defined as the $d$-dimensional 
timelike hyperboloid 
\begin{equation}
\label{hyper}
-(X^0)^2 + \sum_{i=1}^d (X^i)^2 =\ell^2~,
\end{equation}
where $\ell$ is the ${\rm dS}_d$ radius, embedded into $(d+1)$-dimensional Minkowski space $\mathcal M^{1,d}$, with coordinates $(X^0, X^i)$, $i=1,..., d$, and flat metric 
\begin{equation}
\eta =-(dX^0)^2 + \sum_{i=1}^d (dX^i)^2~.
\end{equation}
The ${\rm dS}_d$ hyperboloid~\eqref{hyper} has the topology of 
$\mathbb R\times S^{d-1}$ and manifest $O(d,1)$ symmetries. 
The~${\rm dS}_d$ metric is the induced metric from $\eta$ on the hyperboloid~\eqref{hyper}.
Henceforth we will restrict to the case of $d=4$. 

\paragraph{Global and conformal coordinates.}
The ${\rm dS}_4$ hyperboloid can be foliated by 3-spheres through the parametrization
\begin{equation}
\label{global}
X^0=\ell \sinh (T/\ell)~, \quad 
X^i= \ell \cosh(T/\ell) y^i\ ,\quad i=1,\dots,4~, 
\end{equation}
where $-\infty<T<\infty$ and $y^i$ parametrize the unit 3-sphere, \emph{viz.} $\sum_{i=1}^4(y^i)^2=1$. 
This choice yields a globally defined set of coordinates 
on ${\rm dS}_4$, with induced metric
\begin{equation}\label{globalmetric}
ds^2 = -dT^2 + \ell^2 \cosh^2(T/\ell)\,d\Omega^2_{3}~,
\end{equation}
where $d\Omega^2_{3}$ is the metric on the unit 3-sphere.
The latter can be further folliated by unit 2-spheres, with line element $d\Omega^2_{2}$, as
\begin{equation}
d\Omega^2_{3}= d\Theta^2 
+ \sin^2\Theta\, d\Omega^2_{2}~,
\end{equation}
where $0\leq\Theta\leq\pi$ and the points $\Theta=0, \pi$
are conventionally refer to as the north and south poles of
the (global) 3-sphere.

The causal structure of ${\rm dS}_4$ can be exhibited by going to  
conformal coordinates, defined in terms of the conformal time 
$\tau$, as
\begin{equation}
\tan\frac{\tau}2 :=\tanh\frac{T}{2\ell}~, 
\end{equation}
where thus $-\pi/2<\tau<\pi/2$, and such that
\begin{equation}
ds^2=\frac{-d\tau^2+d\Omega^2_{3}}{\cos^2 \tau}  ~.
\end{equation}
The conformal boundaries
\begin{equation}
\mathcal I^{\pm}:=\tau^{-1}(\pm\pi/2)~,
\end{equation}
are termed the past and future null infinities, $\mathcal I^-$
and $\mathcal I^+$, respectively, both having a $S^3$ topology. 
A null geodesic originated at the north (south) pole in the 
infinite past $\mathcal I^-$ reaches the south (north) pole 
in the infinite future $\mathcal I^+$, as depicted in the 
Penrose diagram of Figure~\hyperlink{Fig:1}1 below.
\vspace{0.5cm}
\begin{center}
\begin{tikzpicture}
\node at (-3.5,3.3) {$\mathcal I^+$};
\node at (-3.5,-0.3) {$\mathcal I^-$};
\node[right, rotate=90] at (-5.3,0.5) {\footnotesize North pole};
\node[right, rotate=-90] at (-1.7,2.5) {\footnotesize South pole}; 
\draw [thick] (-5,0)--(-5,3)--(-2,3)--(-2,0)--(-5,0);
\draw[thick, dashed] (-5,0)--(-2,3);
\draw[thick, dashed] (-5,3)--(-2,0);
\fill[fill=blue!5] (1,0)--(4,0)--(4,3);
\draw [thick] (1,0)--(1,3)--(4,3)--(4,0)--(1,0);
\draw[thick, dashed] (1,0)--(4,3);
\node at (3.4,0.9) {$\mathcal O^-$};
\node at (2.4,1.8) {$\mathcal H^-$};
\fill[fill=blue!5] (5,3)--(8,0)--(8,3);
\draw [thick] (5,0)--(5,3)--(8,3)--(8,0)--(5,0);
\draw[thick, dashed] (5,3)--(8,0);
\node at (7.3,2.2) {$\mathcal O^+$};
\node at (6.3,1.2) {$\mathcal H^+$};
\node at (2.5,3.3) {$\mathcal I^+$};
\node at (2.5,-0.3) {$\mathcal I^-$};
\node at (6.5,3.3) {$\mathcal I^+$};
\node at (6.5,-0.3) {$\mathcal I^-$};
\node [darkred] at (4,1.5) {\textbullet};
\node at (4.5,1.5) {$\mathcal O_S$};
\node [darkred] at (8,1.5) {\textbullet};
\node at (8.5,1.5) {$\mathcal O_S$};
%
\node[text width=16cm, text justified] at (1.5,-2.7){
\small {\hypertarget{Fig:1} \bf Fig.~1}: 
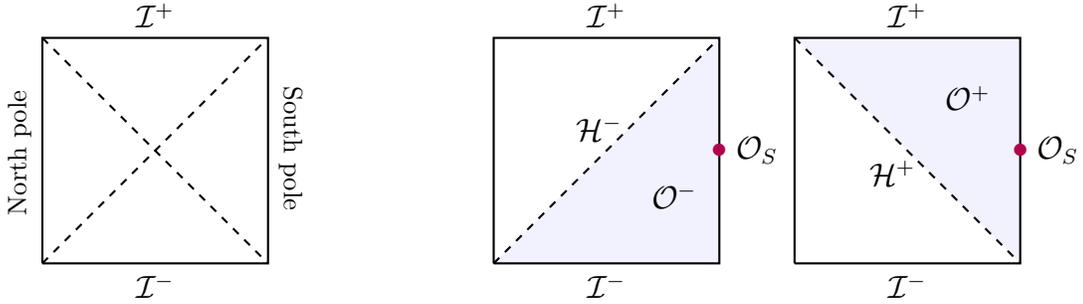
\sffamily{Penrose diagram of ${\rm dS}_4$. 
At the left side, the vertical axis is the conformal time 
$\tau\in [-\pi/2,+\pi/2]$, with the past and future null 
infinities $\mathcal I^-$ and $\mathcal I^+$ as conformal 
boundaries. 
The horizontal axis is the polar coordinate $\Theta\in[0,\pi]$ 
of the global $S^3$, with north and south poles defined by
the points $\Theta=0, \pi$, respectively.
The right hand side displays the causal past $\mathcal O^-$
and causal future $\mathcal O^+$ of an observer at the south 
pole, where $\mathcal H^-$ and $\mathcal H^+$ denote its past 
and future horizons.
}};
\end{tikzpicture}
\end{center}

\paragraph{Static coordinates, inertial observers 
and Rindler wedges.}
An inertial observer $\mathcal O_S$ located at the south pole 
of the global 3-sphere is causally connected to its Rindler wedge 
\begin{equation}
\mathfrak R_S :=\mathcal O^- \cap \mathcal O^+~,
\end{equation}
where $\mathcal O^-$ and $\mathcal O^+$ are the causal 
past and future of $\mathcal O_S$, as indicated in
Figure~\hyperlink{Fig:1}1.
The Rindler wedge $\mathfrak R_S$ represents the set of all points 
in ${\rm dS}_4$ that can send signals to and receive signals from 
$\mathcal O_S$, and its boundary defines an observer-dependent 
cosmological horizon that surrounds $\mathcal O_S$
\begin{equation}
\mathcal H :=\partial\,\mathfrak R_S~.
\end{equation}

Taking $\mathcal O_S$ to follow the worldline 
\begin{equation}
X^1(\mathcal O_S)=\sqrt{\ell^2 + (X^0(\mathcal O_S))^2}~,\quad 
X^i(\mathcal O_S)=0~,\quad i=2,3,4~,
\end{equation}
its Rindler wedge can be parameterized as
\begin{equation}\label{static}
X^0= \sqrt{\ell^2- \hat r^2} \sinh(\hat t/\ell)~, \quad
X^1=\sqrt{\ell^2- \hat r^2} \cosh(\hat t/\ell)~,\quad 
X^i = \hat r \hat y^i ~,\quad  i=2,3,4\ ,
\end{equation}
where $-\infty<\hat t<\infty$, $0\leqslant \hat r <\ell$, 
and $\sum_{i=2}^4 (\hat y^i)^2=1$. The above parametrization 
yields the line element
\begin{equation}
\label{staticmetric}
ds^2= - \left(1-\frac{\hat r^2}{\ell^2}\right) d\hat t^2 
+ \frac{d\hat r^2}{1 -  \frac{\hat r^2}{\ell^2}}    
+ \hat r^2 d\hat\Omega^2_2\ .
\end{equation}
The metric~\eqref{staticmetric} thus describes the worldline of 
the observer $\mathcal O_S$, located at $\hat r=0$, surrounded by a cosmological horizon $\mathcal H$, placed at $\hat r=\ell$.

Inside $\mathfrak R_S$, there exists a notion of Killing time, 
as follows from the explicit time independence of the metric
\eqref{staticmetric}. The Killing vector $\partial_{\hat t}$ is 
timelike inside $\mathfrak R_S$ and null along the (Killing) 
horizon~$\mathcal H$.
\begin{center}
\begin{tikzpicture}
\fill[fill=blue!10] (3,0)--(1.5,1.5)--(3,3);
\node at (2.4,1.5) {$\mathfrak R_S$};
\draw [thick] (0,3)--node[above]{$\mathbf{\mathcal I^+}$}(3,3);
\draw [thick] (0,0) --node[below] {$\mathbf{\mathcal I^-}$}(3,0); 
\draw [thick] (3,0)--(3,3);
\node [darkred] at (3,1.5) {\textbullet};
\node at (3.5,1.5) {$\mathcal O_S$};
\draw [thick] (0,0) --(0,3);
%
\draw [thick, dashed] (0,0) --(3,3); 
\node at (2,0.6) {$\mathcal H$};
\node at (2,2.4) {$\mathcal H$};
\draw [thick, dashed] (0,3)--(3,0);
\node[text width=8cm, text justified] at (9,1.5){
\small {\hypertarget{Fig:2} \bf Fig.~2}: 
\sffamily{
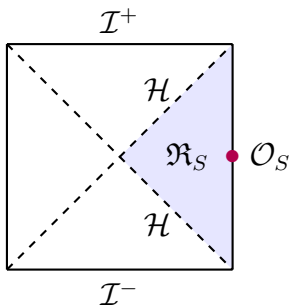
Rindler wedge $\mathfrak R_S=\mathcal O^- \cap \mathcal O^+$
of an observer $\mathcal O_S$ at the south pole.
The cosmological horizon $\mathcal H=\partial\,\mathfrak R_S$ 
is a bifurcated Killing horizon, with bifurcation point the 
intersection $\mathcal H^+\cap\mathcal H^-$ at center of the 
diagram.
}};
\end{tikzpicture}
\end{center}
\paragraph{Temperature and entropy.}
In order to zoom in on the near horizon region, one introduces a dimensionless parameter, say~$\hat \varepsilon\ll1$, as
\begin{equation}
\label{nearhorizon}
\hat r = \ell \Big(1-\frac{\hat\varepsilon^2}{2}\Big)~,
\end{equation}
thus enforcing $\hat r\rightarrow \ell$ as $\hat \varepsilon \rightarrow 0$. 
In this limit, the $(\hat r, \hat t )$ sector of the line element \eqref{staticmetric} is well approximated by the Rindler geometry
\begin{equation}
\label{metric1}
ds^2 \approx  \ell^2 \Big( d\hat\varepsilon^2 
-  \frac{\hat\varepsilon^2}{\ell^2} d\hat t^2\Big) + \cdots
\end{equation}
Periodicity at finite temperature $T_{\rm dS}=\beta^{-1}$ in imaginary time $\hat t \sim \hat t + i\beta$, implies  
\begin{equation}
\label{temperature}
T_{\rm dS} = \frac{1}{2\pi \ell} ~,
\end{equation}
with conjugated Gibbons-Hawking entropy~\cite{Gibbons:1977mu}
\begin{equation}\label{dSS}
{\mathcal S}_{\rm dS} 
= \frac{A_{\mathcal H}}{4G_4}= \frac{\pi\ell^2}{G_4} ~,
\end{equation}
where $A_{\mathcal H}=4\pi\ell^2$ is the area of the cosmological horizon and $G_4$ denotes Newton's constant in four dimensions.

\section{De Sitter space with antipodal defects}\label{Sec:3}
The purpose of this section is twofold. We first introduce an extended set of coordinates that covers both, northern and southern Rindler wedges, and that describes the worldline of two antipodal and causally disconnected observers ${\cal O}_N$ and ${\cal O}_S$.
We then show that, in these coordinates, the $S^2$ geometry of the cosmological horizon admits a family of exact deformations\footnote{By exact deformations we simply mean that the deformed geometry satisfy Einstein's equation.} defined through the orbifold $S^2/\mathbb Z_q$, whose fixed points gives raise to a pair of antipodal minimal surfaces, each one containing the worldline of one observer.
\subsection{Maximally extended coordinates}\label{Sec:3.1}
\paragraph{Coordinates.} 
The hypersurface equation~\eqref{hyper} can be fulfilled by 
parametrizing the embedding coordinates $X \in \mathcal M^{1,4}$ as
\begin{align}
\label{embedding}
X_0 = &\sqrt{\ell^2 - \xi^2}\,\cos\theta\,\sinh(t/\ell)~,\quad
X_1 =\sqrt{\ell^2 - \xi^2}\,\cos\theta\,\cosh(t/\ell)~,  \\ \nonumber
X_2 &= \xi\,\cos\theta ~,\quad
X_3 =\ell \sin\theta \cos\phi~,\quad
X_4 = \ell \sin\theta \sin\phi ~,
\end{align}
where 
\begin{equation}
-\infty<t<\infty~,\quad 
-\ell<\xi<\ell~,\quad 
0\leq \theta\leq\pi~,\quad  
0\leq\phi<2\pi~.
\end{equation}
The resulting line element, that we shall denote by $g_4$, is 
\begin{equation}
\label{metric}
g_4=\ell^2 ( d\theta^2 + \sin^2\theta \, d\phi^2 )
+\cos^2\theta \left[- \left(  1 -  \frac{\xi^2}{\ell^2}\right) dt^2 
+ \frac{d\xi ^2}{1 -  \frac{\xi^2}{\ell^2}} \, \right]~,
\end{equation}
describes the union 
\begin{equation}\label{doubleR}
\mathfrak R_N \cup \mathfrak R_S~,
\end{equation}
of the Rindler wedges of two inertial observers
\begin{equation}\label{ONS}
\mathcal O_N := (\theta=0,~\xi=0)~, \qquad
\mathcal O_S := (\theta=\pi,~\xi=0)~,
\end{equation}
where we have chosen 
\begin{equation}
X^1|_{\mathfrak R_N}>0~, \qquad 
X^1|_{\mathfrak R_S}<0~.
\end{equation}
It follows that
\begin{equation}\label{NSconventions}
\theta\,|_{\mathfrak R_N} \in \Big[0, \frac{\pi}{2}\Big)~,\qquad 
\theta\,|_{\mathfrak R_S} \in \Big(\frac{\pi}{2}, \pi\Big]~,
\end{equation}
while $-\infty<t<\infty$, $-\ell<\xi<\ell$ and $0\leq\phi<2\pi$ in each Rindler wedge.

The northern and southern observers $\mathcal O_N$ and $\mathcal O_S$
(as well as any other pair of points located at opposite hemispheres, \emph{i.e.} with $\theta<\pi/2$ and $\theta>\pi/2$, respectively) are causally disconnected, as a light ray can never cross the equator $\theta=\pi/2$.

The norm of the Killing vector $\partial_t$  
\begin{equation}
\label{killing}
(\partial_t)^2= 
\cos^2\theta \left(\frac{\xi^2}{\ell^2} -1\right)~,
\end{equation} 
is negative inside the two Rindler wedges~\eqref{doubleR}
and null on its boundary at $\xi=\pm \ell$. The latter defines 
the bifurcated Killing horizon $\mathcal H$, which has
a fixed time topology of a 2-sphere of radius~$\ell$. 

\paragraph{Topology and isometries.} 
The line element~\eqref{metric} is a ${\rm dS}^\pm_2$ fibration over $S^2$, where the ${\rm dS}^\pm_2$ factor is radially extended (recall that $-\ell<\xi<\ell$). 
The fibration preserves the manifest $SL(2,\mathbb R)$ isometries 
of ${\rm dS}^\pm_2$, which contains the generator $\partial_t$ of Rindler 
time translations, while it breaks the manifest $SO(3)$ symmetry 
of the $S^2$ factor down to a $U(1)$. The latter is generated by the rotational Killing vector $\partial_\phi$.
Thus, for fixed $\theta~\notin~\big\{0, \frac{\pi}{2}, \pi \big\}$, 
the manifest isometries of~\eqref{metric} form a 
$U(1)\times SL(2,\mathbb R)$. 
For $\theta=0, \pi$, they are broken to $SL(2,\mathbb R)$, 
while only the $U(1)$ factor is retained at the equator 
$\theta=\frac{\pi}{2}$.
%
\paragraph{Comparison with global coordinates.} 

Comparing the parametrizations of the embedding coordinates 
$X^0$ and $X^1$ in equations~\eqref{global} and~\eqref{embedding}, 
we find 
\begin{equation}\label{transf}
\cosh^2(T/\ell)\sin^2\Theta
=\frac{\xi^2}{\ell^2}\cos^2\theta + \sin^2\theta ~, \qquad
\frac{\sinh^2(T/\ell)}{1-\cosh^2(T/\ell)\sin^2\Theta}
= \sinh^2(t/\ell)~.
\end{equation}
From the first equation (and by the choice \eqref{NSconventions}), 
it follows that the worldlines of $\mathcal O_N$ and ${\cal O}_S$ 
are mapped to the north and south poles of the global $S^3$, 
located at $\Theta=0, \pi$, respectively.
From the second equation and by comparing the signs of $X^0$, 
it also follows that
\begin{equation}
t|_{\Theta=0}=T~,\qquad 
t|_{\Theta=\pi}=-T~,
\end{equation}
that is, the local time runs forwards in 
$\mathfrak R_N$ and backwards in $\mathfrak R_S$.

Combining the Killing equation~\eqref{killing} with the first transformation in~\eqref{transf}, we find that the cosmological horizon $\mathcal H$ is sent to the points $(\tau, \Theta)$ of the Penrose diagram satisfying
\begin{equation}
\sin^2\Theta=\cos^2\tau~,
\end{equation}
represented by the two diagonals of the Penrose diagram,   
\emph{viz.}
\begin{equation}
\mathcal H = \partial (\mathfrak R_N \cup \mathfrak R_S)~,
\end{equation}
as displayed in Figure~\hyperlink{Fig:3}3 down below.
\paragraph{Comparison with static coordinates.} 
In order to map the extended coordinates system, that covers
$\mathfrak R_N \cup \mathfrak R_S$, to the standard static coordinates
covering a single wedge, $\mathfrak R_S$, we parametrize the unit 
static 2-sphere in~\eqref{staticmetric} as
\begin{equation}
d\hat\Omega^2_2=d\hat\theta^2 + \sin^2\hat\theta d\hat\phi^2~.
\end{equation}
Comparison between the embedding space coordinates~\eqref{static} 
and~\eqref{embedding} yields the transformations
\begin{equation}\label{map1}
\hat t= {\rm sign}(\cos\theta)\,t~,\qquad
\hat \phi=\phi~,
\end{equation}
and 
\begin{equation}\label{map2}
\tan \hat \theta=\frac{\ell\tan\theta}{\xi}~,\qquad 
\hat r= \sqrt{\ell^2\sin^2\theta+\xi^2\cos^2\theta}~.
\end{equation}
The last two relations can be inverted as follows
\begin{equation}\label{map3}
\sin \theta =\frac{\hat r\sin\hat\theta}{\ell}~,\qquad 
\xi=\frac{\ell\hat r \cos \hat \theta}{\sqrt{\ell^2
-\hat r^2\sin^2\hat \theta}}~.
\end{equation}

Transformations~\eqref{map1}--\eqref{map3} make explicit the map from the regions defined by extended system with $0\leq \theta<\pi/2$ and $\pi/2 <\theta\leqslant \pi$, to
$\mathfrak R_N$ and $\mathfrak R_S$, respectively. 
\vspace{0.5cm}
\begin{center}
\begin{tikzpicture}[scale=0.9]
\fill[fill=yellow!10] (-5,2)-- (-3,2)    
arc (0:180:20mm);                        
\fill[fill=blue!10] (-5,2)-- (-3,2)    
arc (0:-180:20mm); 
\fill[fill=yellow!10] (-3,2) 
arc[start angle=0,end angle=-180, 
x radius=2, y radius=0.5];
\fill[fill=yellow!10] (-5,4)--(-7,2)--(-3,2);
\draw [thick] (-5,2) circle (2cm);
\node [darkred] at (-5,4) {\textbullet};
\node at (-5,4.4) {$\mathcal O_N$};
\node [darkred] at (-5,0) {\textbullet};
\node at (-5,-0.4) {$\mathcal O_S$};
\draw[thick, dashed] (-3,2) 
arc[start angle=0,end angle=180, 
x radius=2, y radius=0.5];
\draw[thick] (-3,2) 
arc[start angle=0,end angle=-180, 
x radius=2, y radius=0.5];
\draw [thick, ->] (-2.5,2) -- (-1.2,2);
\fill[fill=yellow!10] (0,0)--(2,2)--(0,4);
\fill[fill=blue!10] (4,0)--(2,2)--(4,4);
\node [darkred] at (0,2) {\textbullet};
\node [darkred] at (4,2) {\textbullet};
\draw [thick] (0,4) --node[above] {$\mathbf{\mathcal I^+}$}(4,4);
\draw [thick] (0,0) --node[below] {$\mathbf{\mathcal I^-}$}(4,0);                        
\draw [thick] (4,0)--(4,4); 
\node at (4.5,2) {$\mathcal O_S$};
\draw [thick] (0,0) --(0,4);
\node at (-0.5,2) {$\mathcal O_N$};
\node at (1,2) {$\mathfrak R_N$};
\node at (3,2) {$\mathfrak R_S$};
%
\draw [thick, dashed] (0,0) --(4,4); 
\draw [thick, dashed] (0,4)--(4,0);
\node[text width=16cm, text justified] at (-1.3,-3){
\small {\hypertarget{Fig:3} \bf Fig.~3}: 
\sffamily{
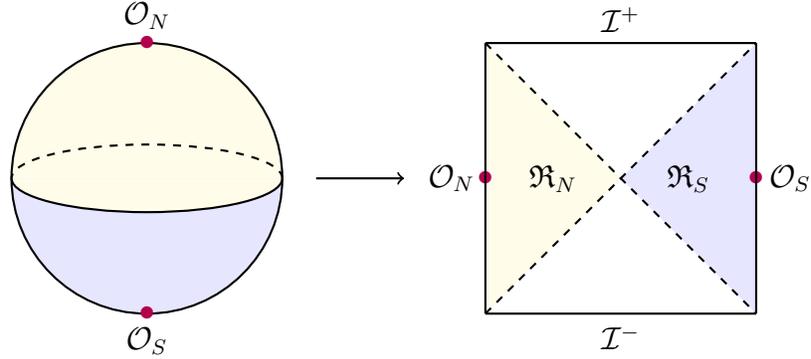
Global depiction of the maximally extended coordinates. 
The worldlines of the observers $\mathcal O_N$ and $\mathcal O_S$ 
are sent to the global north and south poles $\Theta=0, \pi$, 
respectively. 
The north hemisphere of the 2-sphere is mapped to the northern 
Rindler wedge $\mathfrak R_N$. Likewise, the south hemisphere 
is sent to the southern Rindler wedge $\mathfrak R_S$.
}}; 
\end{tikzpicture}
\end{center}
\subsection{De Sitter horizon and Thurston's spindle}
\label{Sec:3.2}
\paragraph{$S^2/\mathbb Z_q$ orbifold and spindle geometry.}
A two-dimensional orbifold~\cite{thurston2014three} $\widehat \Sigma_n=\widehat \Sigma (q_1, ... , q_n)$ is a (closed and orientable) Riemann surface $\Sigma$, possibly endowed with a metric structure, with $n$ marked points $x_i \in \Sigma$, $1\leq i\leq n$, referred to as orbifold points. 
Locally, a neighborhood around an orbifold point $x_i$ is coordinatized by $z_i \in \mathbb C/ \mathbb Z_{q_i}$, where $q_i \in \mathbb Z_{\geq1}$ is an integer termed as anisotropy parameter. 
We thus have the cyclic identification
\begin{equation}
z_i \sim \exp\Big(\frac{2\pi i}{q_i}\Big) z_i ~.
\end{equation}
One can choose a Riemannian metric $g(\widehat \Sigma_n)$ compatible with the orbifold structure such that, in terms of the polar coordinate $z_i=re^{i\phi}$ and locally around an orbifold point $x_i$, one has
\begin{equation}
g = dr^2 + \frac{r^2}{q_i^2} \, d\phi^2~.
\end{equation}
The anisotropy parameters hence induce conical singularities on the metric $g$ at every orbifold point, with deficit angles given by
\begin{equation}
\Delta \phi_i = 2\pi \Big(1-\frac{1}{q_i}\Big)~,
\end{equation}
so that in the limit\footnote{In taking this limit, one assumes the analytic continuation of the anisotropy parameter $q$ to the real numbers.} 
$q_i\rightarrow 1$ the point $x_i$ becomes simply a non-singular marked point. 

Typically and in the simplest cases, an orbifold can be modelled as the quotient $\widehat \Sigma= \Sigma/\Gamma$ of a smooth surface $\Sigma$ by a discrete group $\Gamma$.
A particular family of two-dimensional orbifolds are the \emph{spindle geometries} $S^2(q_1,q_2)$, which in the special case where the two anisotropy parameters are equal $q_1=q_2=q$, can be expressed as the quotient 
\begin{equation}\label{quotient}
S^2(q,q) \cong S^2/\mathbb Z_q~. 
\end{equation}
On the unit $S^2$, considering the angular coordinates $(\theta, \phi)$, where $0\leq \theta \leq \pi$ and $0\leq \phi<2\pi$, with $\mathbb Z_q$-orbifold points at the poles $\theta=0, \pi$, the spindle metric can be taken to be
\begin{equation}\label{spindle}
g_{\rm spindle} = d\theta^2 + \frac{\sin^2 \theta}{b^2} \,d\phi^2\ ,
\end{equation}
where $b=b(\theta)$ is any smooth and positive function with the asymptotic behaviour
\begin{equation}\label{asymp}
b(\theta)=\left\{ 
\begin{array}{cc} 
q + \mathcal O(\theta^2)~, 
&\theta\rightarrow 0~, \\[1mm]
q + \mathcal O((\theta-\pi)^2)~,
&\theta\rightarrow \pi~.
\end{array} \right.
\end{equation}

\paragraph{Orbifolding the horizon.}
Following the rationale just described and by means of the identification $\phi\sim\phi + 2\pi q^{-1}$, the north and south poles of the horizon $\mathcal H \cong S^2$ can be taken to be two $\mathbb Z_q$-orbifold 
points with the same anisotropy parameter $q$.
By taking the quotient $S^2/\mathbb Z_q$, the round geometry of $\mathcal H$ deforms into that of a spindle orbifold\footnote{Strictly speaking, the spindle $S^2(q_1,q_2)$ with $q_1=q_2$ is termed the \emph{football}. 
Nonetheless, since the latter is a particular case of the former, 
we will continue using the term spindle.}. 
As a result, the ${\rm dS}_4$ line element~\eqref{metric} takes the conically singular form
\begin{equation}\label{dS4q}
\widehat g_4= \ell^2 g_{\rm spindle} + w^2 g_2^\pm~,
\end{equation}
where $g_{\rm spindle}$ is the metric on $S^2/\mathbb Z_q$ (defined in~\eqref{spindle}) and $g_2^\pm$ denotes the metric on ${\rm dS}_2^\pm$, \emph{viz}.
\begin{equation}\label{split}
g_{\rm spindle}= d\theta^2 + \frac{\sin^2 \theta}{q^2} \,d\phi^2~,\qquad
g_2^\pm = -\left(1-\frac{\xi^2}{\ell^2}\right) dt^2 
+ \frac{d\xi ^2}{1 -  \frac{\xi^2}{\ell^2}}~.
\end{equation}
In addition, the warp factor $w(\theta)=\cos\theta$ satisfies the holonomy conditions 
\begin{align}\label{bc}
w^2 \big |_{\theta=0, \pi}=1~, \quad 
(w^2)^\prime \big |_{\theta=0, \pi} =0~.
\end{align}
From here and in what follows, we denote the conically singular manifold endowed with the metric~\eqref{dS4q} as 
\begin{equation}\label{dSq}
\widehat{\rm dS}_4:={\rm dS}_4/\mathbb Z_q~.
\end{equation} 

Due to the holonomy conditions~\eqref{bc}, the induced 
two-dimensional geometry at the orbifold points is given by 
\begin{equation}
\label{SigmaNS}
(\Sigma_N, h)
=\big(\widehat{\rm dS}_4, \widehat g_4\big)\big|_{\theta=0}~,\quad
(\Sigma_S, h)
=\big(\widehat{\rm dS}_4, \widehat g_4\big)\big|_{\theta=\pi}~,
\end{equation}
where the induced metric $h=g_2^\pm$. We will refer to the codimension two submanifolds $(\Sigma_N, h)$ and $(\Sigma_S, h)$ defined in~\eqref{SigmaNS} as \emph{defects}.

Locally, the manifolds ${\rm dS}_4$ and $\widehat{\rm dS}_4$ have the same curvature. 
However, globally they differ. The nontrivial holonomy around the conical singularities induced by the anisotropy parameter $q$ contributes with one $\delta$-function singularity to the Riemann curvature tensor~\cite{Fursaev:1995ef} for each orbifold point.  
In turns, the Ricci scalar
\begin{equation}\label{Ricciq}
R^{(q)} = R - \sum_{I=N, S} 4\pi 
\Big(1-\frac{1}{q}\Big) \delta_{\Sigma_I}~,
\end{equation}
where $R= R^{(q)}|_{q=1}$ is the Ricci scalar of the regular manifold ${\rm dS}_4$, 
and $\delta_{\Sigma_I}$ is a projector onto the defects, \emph{viz.} 
$\int_{\widehat{\rm dS}_4} f \delta_{\Sigma_I}=\int_{\Sigma_I} f|_{\Sigma_I}$.
Integrating~\eqref{Ricciq} over the conically singular manifold $\widehat{\rm dS}_4$ yields the total action
\begin{align}\label{Itotal}
I[\widehat{\rm dS}_4] &
:= \frac{1}{16\pi G_4}\int_{\widehat{\rm dS}_4}d^4 x \sqrt{-\widehat g_4}
\Big( R^{(q)}-\frac{6}{\ell^2}\Big)\nonumber\\
&~=\frac{1}{16\pi G_4} 
\int_{\widehat{\rm dS}_4\setminus(\Sigma_N\cup\Sigma_S)}
d^4 x \sqrt{-\widehat g_4}\Big( R- \frac{6}{\ell^2}\Big) -\sum_{I=N, S} 
\mathcal T_q\int_{\Sigma_I} d^2 y\sqrt{-h}~,
\end{align}
which thus consists of a bulk integral, that excludes 
the submanifolds $\Sigma_N$ and $\Sigma_S$, plus one 
copy of the Nambu-Goto action for each defect, coupled 
through the tension
\begin{equation}\label{tension}
\mathcal T_q = \frac{1}{4 G_4} \Big(1-\frac{1}{q}\,\Big)~.
\end{equation} 
This yields localized stress-energy tensors
\begin{equation}\label{Tij}
T_{ij}^{N} = T_{ij}^{S}= \mathcal{T}_q \,h_{ij} ~,
\end{equation}
with support on the defects $\Sigma_N$ and $\Sigma_S$, 
respectively. 
The construction is illustrated in Figure~\hyperlink{Fig:4}4 below.
\begin{center}
\begin{tikzpicture}[scale=0.9]
\draw [thick] (-4,0) circle (2cm);
\node [darkred] at (-4,2) {\textbullet};
\node at (-4,2.3) {$x_N$};
\node [darkred] at (-4,-2) {\textbullet};
\node at (-4,-2.3) {$x_S$};
\node at (-6,2) {$\mathcal H\cong S^2$};
\draw[thick, dashed] (-2,0) 
arc[start angle=0,end angle=180, 
x radius=2, y radius=0.5];
\draw[thick] (-2,0) 
arc[start angle=0,end angle=-180, 
x radius=2, y radius=0.5];
\node at (-0.5,0.4) {$S^2/\mathbb Z_q$};
\draw [thick, ->] (-1.3,0) -- (0.3,0);
\fill[fill=yellow!10] (1,2.3)--(0,1.7)--(5,1.7)--(6,2.3);
\fill[fill=blue!10] (1,-1.7)--(0,-2.3)--(5,-2.3)--(6,-1.7);
\draw [thick] (1,2.3) --(6,2.3);
\draw [thick] (0,1.7) --(5,1.7);
\draw [thick] (0,1.7) --(1,2.3);
\draw [thick] (6,2.3) --(5,1.7);
\coordinate (N) at (3,2);\coordinate (S) at (3,-2);
\node [darkred] at (3,2) {\textbullet};
\node [darkred] at (3,-2) {\textbullet};
\draw[thick] (N)to[out=-20,in=20](S);
\draw[thick] (N)to[out=-150,in=150](S);
\draw[thick, dashed] (4.1,0) 
arc[start angle=0,end angle=180, 
x radius=1.05, y radius=0.3];
\draw[thick] (4.1,0) 
arc[start angle=0,end angle=-180, 
x radius=1.05, y radius=0.3];
\node at (6,1.7) {$(\Sigma_N, h)$};
\node at (0.1,-1.6) {$(\Sigma_S, h)$};
\draw [thick] (1,-1.7) --(6,-1.7);
\draw [thick] (0,-2.3) --(5,-2.3);
\draw [thick] (0,-2.3) --(1,-1.7);
\draw [thick] (6,-1.7) --(5,-2.3);
\draw [thick, ->] (3.05,0) -- (4.05,0);
\node at (5.1,0) {$\ell_q=\ell/q $};
\node[text width=16cm, text justified] at (0,-5) 
{\small {\hypertarget{Fig:4}\bf Fig.~4}: 
\sffamily{
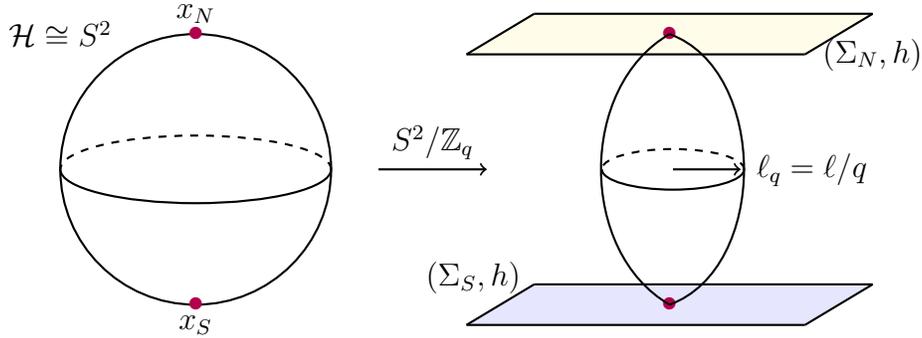
dS horizon with two marked points $x_N$ and $x_S$.
After orbifolding, the spherical geometry globally deforms to 
that of the $S^2/\mathbb Z_q$ spindle, with radius $\ell_q:=\ell/q$. 
The conical singularities are resolved by coupling two copies 
of the Nambu-Goto action to the Einstein--Hilbert 
bulk action, one for each orbifold point and with 
fixed induced metric $h=g_2^\pm$.}
};
\end{tikzpicture}
\end{center}

\section{De Sitter entropy from entanglement}
\label{Sec:4}
In this section, we first argue that the entropy of four-dimensional dS space follows from the entanglement between the two disconnected conformal boundaries $\mathcal I^-$ and $\mathcal I^+$.
We then consider the entanglement between the two causally disconnected Rindler wedges ~$\mathfrak R_S$ and~$\mathfrak R_N$, and show that the corresponding entanglement entropy obeys an area law in terms of the minimal surfaces that arise as the set of fixed points of the bulk $\mathbb Z_q$ action.

\subsection{Entanglement between disconnected boundaries}
\label{Sec:4.1}
\paragraph{De Sitter holography.} 
Despite the lack of an open/closed string-like duality in dS space,
it has been argued~\cite{Witten:2001kn, Strominger:2001pn, Park:1998qk, Hull:1998aa, Banados:1998tb, Park:1998yw, Bousso:1999xy, Bousso:1999cb, Bousso:2000nf, Balasubramanian:2001aa, Klemm:2001ea, Cacciatori:2001un, Balasubramanian:2002aa, Bousso:2002aa, Spradlin:2001nb, Balasubramanian:2002zh, Maldacena:2002vr, Anninos:2011ui} that the rationale underlying the formulation of the
AdS/CFT correspondence may somehow be extendable to positively 
curved backgrounds, wherein quantum gravity on dS space should be 
dual to an Euclidean, possibly non-unitary, conformal field theory defined 
on (or embedded in) the spacelike conformal boundaries $\mathcal I^{\pm}$.
However, the current status of this duality is not completely clear and so far only two concrete examples of it have been found, namely the case of three bulk dimensions~\cite{Cacciatori:2001un}---whose dual is an Euclidean Liouville theory---and the higher spin realization spelled out in~\cite{Anninos:2011ui}.

Here, we follow Bousso's approach to holography
\cite{Bousso:1999xy, Bousso:1999cb, Bousso:2000nf} and  
consider the past and future conformal boundaries $\mathcal I^{\pm}$ 
of dS space as two \emph{holographic screens}.
On general grounds, these are special hypersurfaces embedded on
the boundary of a given spacetime which store all the bulk information 
of a manifold with boundaries, with a density of no more than one degree 
of freedom (or one bit of information) per Planck area.
In the case of dS space, based on the observation that any null geodesic 
begins at some point $p^-\in\mathcal I^-$ and ends at some point 
$p^+\in\mathcal I^+$,  Bousso argued that 
an inertial observer in dS space can be characterized in terms  of the pair 
of points $(p^-, p^+ )$, and that half of the bulk region can be \emph{holographically projected} along light rays onto the past infinity $\mathcal I^-$, and likewise the 
other half can be sent to the future infinity~$\mathcal I^+$. 

Relying on the previous arguments, hereafter we shall adopt 
an holographic scheme based on the following working 
hypotheses:
\begin{itemize}
\item[$\diamond$] There exists an holographic duality between
quantum gravity on dS space and two copies\footnote{
This assumption is compatible with Strominger's formulation of the 
dS/CFT correspondence~\cite{Strominger:2001pn}. The causal connection relating 
points on the boundary past sphere to antipodal points on the boundary future
sphere induces an antipodal map which breaks the two copies 
of the conformal group (one copy for each boundary) down to a single 
copy. Hence, there truly exists a \emph{single} dual CFT, localized on 
a \emph{single} sphere.}
of a certain conformal field theory, one copy for each boundary, 
such that the Hilbert space corresponding to the bulk Rindler wedge 
$\mathfrak R_S$ is equivalent to the tensor product
of the past and future CFTs Hilbert spaces, \emph{viz.} 
\begin{equation}\label{holo1}
\mathcal H_{\mathfrak R_S} = \mathcal H_{\mathcal I^+} 
\otimes \mathcal H_{\mathcal I^-} \,,
\end{equation}
and such that the partition functions
\begin{equation}\label{holo2}
\mathcal Z_{\rm QG}[\mathfrak R_S] =
\mathcal Z_{\rm CFT}[S^3]
 \times\mathcal Z_{\rm CFT}[S^3] \,,
\end{equation}
where the right hand side comprises two identical copies of the partition function on the boundary 3-sphere $S^3\cong \mathcal I^\pm$. Furthermore, we assume the existence of an infrared limit in which the bulk quantum gravity theory admits a low energy description in terms of Einstein gravity, where the partition function $\mathcal Z_{\rm QG}$ is well approximated by the Euclidean on-shell gravity action as
\begin{equation} \label{saddle}
\mathcal Z_{\rm QG}[\mathfrak R_S]\approx
\exp\big(-I_{E}[\mathfrak R_S]\big) \,. 
\end{equation}
(In the above, we have used the notation ``$\approx$" to indicate on-shell equality.)

In this limit, the correspondence \eqref{holo2} becomes\footnote{Note that, using the fact that the Euclidean on-shell gravity action equals minus the Gibbons--Hawking entropy, \emph{viz}. $I_{E}\approx - \mathcal S_{dS}$, it follows from~\eqref{holo3} that $\mathcal S_{dS}=2\log \mathcal Z_{\rm CFT}[S^3]$; a relation of this type has been proposed in~\cite{Maldacena:2011mk}~(see also~\cite{Anninos:2012qw}).}
\begin{equation}\label{holo3} 
\exp\big(-\tfrac 12 I_{E}[\mathfrak R_S]\big)\approx
\mathcal Z_{\rm CFT}[S^3]\,.
\end{equation}
\item[$\diamond$] The bulk quantum state $\lvert \mathcal O_S \rangle$ associated to a static observer $\mathcal O_S\in\mathfrak R_S$ can be holographically described by the thermofield double state\footnote{This hypothesis is consistent with Maldacena's reformulation of the dS/CFT correspondence~\cite{Maldacena:2002vr} in terms of the Hartle--Hawking wave function, \emph{viz}. $\Psi_{dS}=\mathcal Z_{\rm CFT}$, whereby bulk correlators are computed using the probability measure $\vert \Psi_{dS}\vert^2=\Psi_{dS}^\ast \Psi_{dS}$. This, in turn, suggests an holographic picture that comprises two copies of the boundary field theory.  Relying on the above argument, the possibility of an holographic duality between dS space and a state of the thermofield double type was first discussed in~\cite{Narayan:2017xca, Jatkar:2018aa}. We thank K.~Narayan for clarifying this point to us.}
\begin{equation}\label{TFD}
\lvert \mathcal O_S \rangle \sim \sum_n e^{-\beta E_n/2}
\lvert{n}\rangle_{\mathcal I^-}\otimes\lvert{n}\rangle_{\mathcal I^+}\,,
\end{equation}
with $\lvert{n}\rangle_{\mathcal I^\pm}\in\mathcal H_{\mathcal I^\pm}$, and where the  boundary modular Hamiltonian $H\lvert n \rangle_{\mathcal I^{\pm}}= E_n\lvert n \rangle_{\mathcal I^{\pm}}$. Consequently, the density matrix built up from the observer state~\eqref{TFD} is thermal 
\begin{equation}\label{Orho}
\rho := \lvert \mathcal O_S \rangle \langle \mathcal O_S \vert 
\sim e^{-\beta H}\,, 
\end{equation}
and it thus encodes all the thermodynamic properties of dS space.
\end{itemize}

As we shall see next, the set of equations \eqref{holo1}--\eqref{Orho} provides
an holographic framework from which the Gibbons--Hawking entropy of dS space can be derived from the entanglement between the two conformal boundaries $\mathcal I^-$ and $\mathcal I^+$.   

\paragraph{Holographic entanglement entropy.}
To begin with, let us recall some basic aspects of entanglement in quantum field theory (for a detailed review see, for instance, \cite{Rangamani:2016dms}). Denoting by $\mathcal B$ the Lorentzian manifold on which the field theory is defined, we pick a codimension one spacelike Cauchy slice $\mathcal C$ (on which we have a state of the system, say $\rho_{\mathcal C}$) which in turns we partition as $\mathcal C= \mathcal A \cup \mathcal A^c$, and such that the Hilbert space of the theory factorizes as
\begin{equation}
\mathcal H = \mathcal H_{\mathcal A} \otimes \mathcal H_{\mathcal A^c}.
\end{equation} 
Importantly, the temporal evolution of the initial data on $\mathcal C$ amounts to reconstruct the state of the theory on the full manifold $\mathcal B$, \emph{viz}.
\begin{equation}\label{BfromC}
\mathcal B = D^+[\mathcal C] \cup D^-[\mathcal C]\,, 
\end{equation}
where $D^\pm[\mathcal C]$ denotes future and past domain of dependence of the Cauchy slice $\mathcal C$.

Within the above setup, the amount of entanglement between the two complementary subsystems $\mathcal A$ and $\mathcal A^c$ is encoded in the entanglement entropy
\begin{equation}\label{EE}
\mathcal S_{\rm E}= 
- {\rm Tr}_{\mathcal A}\,(\hat\rho_{\mathcal A}\log\hat\rho_{\mathcal A})\,,
\end{equation}
where the reduced density matrix $\hat\rho_{\mathcal A}:={\rm Tr}_{\mathcal A^c}(\hat\rho_{\mathcal C})$, with $\hat\rho_{\mathcal C}$ denoting the density matrix 
of the total system $\mathcal C=\mathcal A \cup \mathcal A^c$ (we use the normalization ${\rm Tr}_{\mathcal A}\,\hat\rho_{\mathcal A}=1$). The entropy~\eqref{EE} can be alternatively obtained as the limit
\begin{equation}\label{Renyi}
\mathcal S_{\rm E}
= \lim_{q\to1}\frac{\log {\rm Tr}_{\mathcal A}\,\hat\rho_{\mathcal A}^{\,q}}{1-q}
= -\partial_q \log {\rm Tr}_{\mathcal A}\,\hat\rho_{\mathcal A}^{\,q}\big|_{q=1} ~,
\end{equation}
where the standard approach to compute the $q$-th power of the reduced density matrix is the replica method~\cite{0305-4608-5-5-017}. This prescription is based on the construction of a $q$-fold branched cover, say~$\mathcal B_q$, 
of the manifold $\mathcal B=:\mathcal B_1$ wherein the field theory is defined.
In order to build up $\mathcal B_q$, one simply needs to 
replicate $q$ times the original manifold $\mathcal B$ 
along the entangling region $\mathcal A$ 
(whose entanglement entropy is being computed). 
Due to the cyclicity of the trace operation, the branched cover 
$\mathcal B_q$ naturally admits a $\mathbb Z_q$ action, whose 
fixed points are the boundary of the entangling region 
$\partial \mathcal A$, referred to as the entangling surface. Consequently, the trace of $\hat\rho_{\mathcal A}^{\,q}$ in~\eqref{Renyi} can be computed in terms of the field theory partition function on the branched cover~$\mathcal B_q$ by means of the formula~\cite{0305-4608-5-5-017,Calabrese:2004eu}
\begin{equation}\label{rhoZ}
{\rm Tr}\,\hat\rho_{\mathcal A}^{\, q} = 
\frac{\mathcal Z_{\rm CFT}[\mathcal B_q]}
{(\mathcal Z_{\rm CFT}[\mathcal B_1])^q}~.
\end{equation}

Now we wish to argue that the rationale outlined above can be adapted to the disconnected dS$_4$ boundaries. There exists, however, an important difference: The two conformal boundaries $\mathcal I^+$ and $\mathcal I^-$ are Euclidean manifolds and hence it is not possible to define the temporal evolution of a Cauchy slice containing initial data. Nonetheless, as we shall now propose, it is plausible to define a codimension one Cauchy-like surface, say $\mathcal C_{\Phi}$, in such a way that the state of the dual theory on the full boundary $\mathcal B$ can be reconstructed \emph{via modular evolution} of the data on $\mathcal C_{\Phi}$. 

To this end, using the fact that null geodesics induce the antipodal map\footnote{
By considering the embedding $S^d\hookrightarrow \mathbb R^{d+1}$, 
with embedding coordinates $X^0,..., X^d\in\mathbb R^{d+1}$, the antipodal 
map $\pi : \mathbb R^{d+1} \longrightarrow \mathbb R^{d+1}$ is defined as the composition of $d+1$ reflections, each acting on one coordinate as 
$(X^0, X^1,..., X^d) \stackrel{\pi}\longmapsto (-X^0, -X^1,..., -X^d)$.}
\cite{Witten:2001kn, Strominger:2001pn}
\begin{equation}\label{anti}
\pi : S^3 \rightarrow S^3\,,
\end{equation}
that sends every point on the past 3-sphere $\mathcal I^-$ 
to an antipodal point on the future 3-sphere $\mathcal I^+$, 
we take the boundary manifold $\mathcal B$ to be defined as
\begin{equation}\label{Bpi}
\mathcal B:= (\mathcal I^{-}\cup\mathcal I^{+} )~\big/~\pi \,.
\end{equation}
It follows that $\mathcal B$ has the topology of a \emph{single} 3-sphere
\begin{equation}\label{S3bdry}
\mathcal B \cong S^3\,,
\end{equation} 
which thus admits an obvious $\mathbb Z_q$ action---given by discrete azimutal identifications---that can be naturally treated as boundary replica symmetry.
Indeed, taking the metric on the boundary 3-sphere to be
\begin{equation}\label{Bmetric}
g_{\mathcal B} =  d\psi^2 
+ \sin^2\psi \big( d\chi^2 + \sin^2\chi\,d\Phi^2\big)\,,
\end{equation}
the discrete $\mathbb Z_q$ identification
\begin{equation}\label{Phi}
\Phi \sim \Phi + \frac{2\pi}{q} ~, ~~ q>1\,,
\end{equation}
leaves invariant the points on the circle $S^1:=\{\chi= 0\}\cup\{\chi=\pi\}$, where each value of $\chi$ contributes with half of the circle;  this can be seen by noting that for every constant value of the angle $\psi$ in~\eqref{Bmetric}, the resulting 2-sphere has two fixed points under the $\mathbb Z_q$ action (its corresponding north and south poles, given precisely by~$\chi= 0, \pi$). The continuous collection of all such points gives raise to the $S^1$ set of fixed points on~$\mathcal B\cong S^3$. 

Next, we choose the Cauchy-like surface $\mathcal C_\Phi$ as the 2-sphere that results from gluing together two disks whose boundaries coincide with the $S^1$ set fixed points of the $\mathbb Z_q$ action on $\mathcal B\cong S^3$, \emph{viz}.
\begin{equation}\label{Clike}
\mathcal C_\Phi = D^2_+ \cup  D^2_-\,, \quad
\partial  D^2_\pm = S^1\,.
\end{equation}
The two disks $D^2_\pm$ can separately be obtained by first acting with the antipodal map~\eqref{anti} on~$\mathcal I^\pm$ and then fixing the azimutal angle $\Phi$ to a constant value. In this way, we interpret the construction of the Cauchy-like slice~\eqref{Clike} as the continuation to Euclidean space of the usual choice of a Cauchy slice at zero time. Moreover, as it is known for the case of spherical entangling regions~\cite{Casini:2011kv}, there exists an explicit expression for the (local) modular Hamiltonian whose analytic continuation generates a $U(1)$ symmetry along Euclidean time. Hence, modular evolution of the data in $\mathcal C_\Phi$ arguably reconstruct the state on the full boundary $\mathcal B$. The complete maneuver is illustrated in Figure~\hyperlink{Fig:5}5 below.
\vspace{0.2cm}
\begin{center}
\begin{tikzpicture}
\draw[thick] (-6,1)arc[start angle=90, end angle=270, x radius=1, y radius=1];
\draw[thick, darkred] (-6,1) 
arc[start angle=90, end angle=-90, x radius=0.2, y radius=1];
\draw[thick, darkred] (-6,1) 
arc[start angle=90,end angle=270, x radius=0.2, y radius=1];
\draw[thick] (-5,1)arc[start angle=90, end angle=-90, x radius=1, y radius=1];
\draw[thick, darkred] (-5,1) 
arc[start angle=90, end angle=-90, x radius=0.2, y radius=1];
\draw[thick, darkred] (-5,1) 
arc[start angle=90,end angle=270, x radius=0.2, y radius=1];
\node at (-7,1.1) {$D^2_+$};
\node at (-3.9,1.1) {$D^2_-$};
\node at (-5.5,-1.5) {$S^1=\partial D^2_\pm$};
\node at (-2.5,0.3) {\footnotesize Gluing};
\draw [thick, ->] (-3,0)--(-2,0);
\draw [thick] (0,0) circle (1cm);
\draw[thick, darkred, dashed] (0,1) 
arc[start angle=90, end angle=-90, 
x radius=0.2, y radius=1];
\draw[thick, darkred] (0,1) 
arc[start angle=90,end angle=270, 
x radius=0.2, y radius=1];
\node at (0,-1.5) {$\mathcal C_\Phi= D^2_+ \cup D^2_-$};
\node at (0,1.3) {\footnotesize\color{darkred}Fixed points of $\mathcal B/\mathbb Z_q$};
\node at (2.3,0.3) {\footnotesize Projecting};
\draw [thick, ->] (1.7,0)--(2.7,0);
\draw [thick] (3.5,-1.5)--(3.5,1.5)--(7.5,1.5)--(7.5,-1.5)--(3.5,-1.5);
\draw [thick, darkred] (5.5,0) circle (1cm);
\node at (5.5,-1.25) {\footnotesize\color{darkred}Entangling surface};
\node at (5.5,0) {\footnotesize$D^2_+$};
\node at (6.9,0.5) {\footnotesize$D^2_-$};
\node at (3.9,1.1) {\footnotesize $\mathcal C_\Phi$};
\node[text width=16cm, text justified] at (0,-4){
\small {\hypertarget{Fig:5}\bf Fig.~5}:
\sffamily{
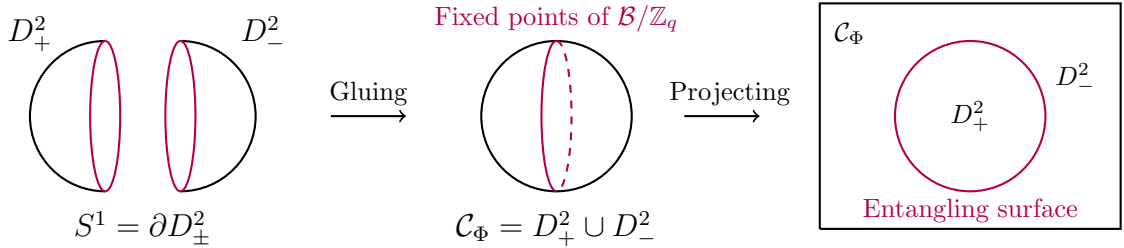
Construction of Cauchy-like slice $\mathcal C_\Phi\cong S^2$. Due to the
existence of the antipodal map~\eqref{anti}, the total boundary
$\mathcal I^-\cup \mathcal I^+$ effectively reduces to a single 3-sphere,
$\mathcal B:= (\mathcal I^{-}\cup\mathcal I^{+} )~\big/~\pi$.  
The latter has a natural replica symmetry whose $S^1$-set fixed points 
coincides with the boundary of the two disks  $D^2_\pm$ that build up
$\mathcal C_\Phi$. Stereographic projection onto the plane exhibits the bipartition
$\mathcal C_\Phi =  D^2_+ \cup D^2_-$ explicitely.
}}; 
\end{tikzpicture}
\end{center}

Having constructed the boundary manifold $\mathcal B$ and the bipartition of the Cauchy-like slice $\mathcal C_\Phi = D^2_+ \cup D^2_-$, we now proceed to compute the entanglement entropy between the two subsystems~$D^2_+$ and $D^2_-$. From the replica formula~\eqref{rhoZ} specialized to $\mathcal A= D^2_+$ 
\begin{equation}\label{rhoZ}
{\rm Tr}\,\hat\rho_{\mathcal \mathcal D^2_+}^{\, q} = 
\frac{\mathcal Z_{\rm CFT}[\mathcal B_q]}
{(\mathcal Z_{\rm CFT}[\mathcal B_1])^q}\,,
\end{equation}
where $\mathcal B_q$ denotes the branched cover of $\mathcal B$, it follows that 
\begin{equation}\label{Sq}
\log {\rm Tr}\,\hat\rho_{D^2_+}^{\,q}
= \log \mathcal Z_{\rm CFT}[\mathcal B_q]
- q\log\mathcal Z_{\rm CFT}[\mathcal B_1]\,.
\end{equation}
Next, using the holographic relation~\eqref{holo3} specialized to the boundary 3-sphere $\mathcal B\cong S^3$, we obtain
\begin{equation}\label{SI}
\log {\rm Tr}\,\hat\rho_{D^2_+}^{\,q}
\approx -\frac 12 I_{E}[\mathfrak R_{S, q}] + \frac q2 I_{E}[\mathfrak R_S]\,.
\end{equation}
In the last expression, $\mathfrak R_{S, q}$ denotes the branched cover of the southern Rindler wedge $\mathfrak R_S=:\mathfrak R_{S, 1}$, which is assumed to inherit the boundary replica symmetry~\cite{Lewkowycz:2013nqa}. Using the locality of the gravity action, we can futher write
\begin{equation}
I_{E}[\mathfrak R_{S, q}] = q I_{E}[\mathfrak R_S/\mathbb Z_q]\,,
\end{equation}
and hence Equation~\eqref{SI} reads
\begin{equation}\label{SIq}
\log {\rm Tr}\,\hat\rho_{D^2_+}^{\,q}
\approx -\frac{q}{2} \Big(I_E[\mathfrak R_S/\mathbb Z_q] - I_E[\mathfrak R_S]\Big)\,.
\end{equation}

In the above, the on-shell value of the gravitational action is obtained from the total action~\eqref{Itotal} restricted to the domain of the southern Rindler wedge $\mathfrak R_S$. In this domain, the bulk $\mathbb Z_q$ action induces a single defect.
The calculation follows from considering the action~\eqref{Itotal} and the line element~\eqref{dS4q} (with the induced metric $h=g_2^\pm$ given in~\eqref{split}): 
\begin{align}\label{IEtotal}
I_{E}[\mathfrak R_S/\mathbb Z_q] &= I[\widehat{dS}_4]\Big|_{\mathfrak R_S}\cr
&=-\frac{1}{16\pi G_4} 
\int_{\widehat{dS}_4\setminus \Sigma_S}
d^4 x \sqrt{g}\Big( R- \frac{6}{\ell^2}\Big)
+\mathcal T_q
\int_{\Sigma_S} d^2 y\sqrt{h}\nonumber\\
&\approx
-\frac{3}{4qG_4} A_{\Sigma_S} \int_{\pi/2}^{\pi} 
d\theta\sin\theta\cos^2 \theta
~+\frac{1}{4G_4}\Big(1-\frac{1}{q}\Big)A_{\Sigma_S}~.
\end{align}
Here, since we are only considering the southern Rindler wedge $\mathfrak R_S$, the integral in $\theta$ should be taken in the domain $\theta\in(\pi/2, \pi]$ (as indicated in~\eqref{NSconventions}), and its value is $1/3$. We have also made use of the value of the tension~\eqref{tension} and  the Einstein's equations $\ell^2 R_{\mu\nu}=3g_{\mu\nu}$.  
Finally, we have denoted the area of $\Sigma_S$ by $A_{\Sigma_S}$; this is readily computed by noting that the Euclidean defect, endowed with the radially extended metric $h=g_2^\pm$, has topology of a 2-sphere of radius $\ell$ (built up from one Euclidean ${\rm dS}_2$ disk for $\xi>0$ and another one for $\xi<0$). Thus, $A_{\Sigma_S}=4\pi\ell^2$, and the on-shell value of the gravity action~\eqref{IEtotal} becomes\footnote{Note that, in the tensionless limit $q\to1$, one recovers the known result in which the on-shell gravity action equals minus the Gibbons--Hawking entropy $I_{E}\approx - \mathcal S_{dS}$.}
\begin{equation}\label{Ionshell2}
I_{E}[\mathfrak R_S/\mathbb Z_q] 
\approx \Big(1-\frac{2}{q}\Big)\frac{A_{\Sigma_S}}{4G_4}
=\Big(1-\frac{2}{q}\Big)\frac{\pi\ell^2}{G_4}\,.
\end{equation}
Using this value in Equation~\eqref{SIq}, we have that
\begin{equation}
\log {\rm Tr}\,\hat\rho_{D^2_+}^{\,q} = (1-q)\, \frac{A_{\Sigma_S}}{4G_4}\,,
\end{equation}
and therefore one finds that the entanglement entropy~\eqref{Renyi} is given by one quarter of the area of the minimal surface $\Sigma_S$:
\begin{equation}
\mathcal S_{\rm E}=\frac{A_{\Sigma_S}}{4G_4}=\frac{\pi\ell^2}{G_4}\,.
\end{equation}
This result reproduces the Gibbons--Hawking entropy~\eqref{dSS}.

\subsection{Entanglement between disconnected bulk regions}
The two causally disconnected Rindler wedges $\mathfrak R_N$ and $\mathfrak R_S$ are naturally entangled. As we shall now see, this bulk notion of entanglement can be understood with no need of holography. 
We begin our bulk analysis by observing that the maximally extended 
coordinates devised in Section~\ref{Sec:2} made manifest the existence
of a bulk antipodal map which sends every point 
inside the (properly Euclideanized) northern Rindler wedge $\mathfrak R_N$
to an antipodal point on $\mathfrak R_S$ (and viceversa).  Thus, when
modded out, the bulk antipodal map enables to describe the bulk topology 
in terms of a \emph{single} 4-sphere.
Moreover, when restricted to a 3-sphere, the bulk antipodal map induces a smaller antipodal map that we interpret as the boundary antipodal map~\eqref{anti} used previously to study the entanglement between the two disconnected dS boundaries.

\paragraph{Bulk antipodal map.}
Upon Wick rotation, the two Rindler wedges $\mathfrak R_N$
and $\mathfrak R_S$ acquires separately the topology of a 
4-sphere\footnote{We recall that an Euclidean Rindler wedge 
has the topology of a single 4-sphere of volume ${\rm Vol}(S^4)$. 
This 4-sphere is the same 
that the one obtained by Wick rotation of the global metric~\eqref{globalmetric}. 
When considering the extended coordinates~\eqref{metric}, 
direct calculation shows that
$$ \int_{\mathfrak R_N\cup\mathfrak R_S} 
d^4x \sqrt{g}= 2\,{\rm Vol}(S^4)~,$$
where the integration over each Rindler wedge contributes 
with a factor of one to the total volume above.
}. 
The Euclidean bulk topology is hence that of  
\begin{equation}
\big(\mathfrak R_N \cup \mathfrak R_S\big)_E 
\cong S^4_{\mathfrak R_N} \cup S^4_{\mathfrak R_S} ~.
\end{equation}
Using the Euclideanized embedding coordinates introduced
in~\eqref{embedding}, it is possible to define the invertible antipodal 
map 
\begin{equation}\label{Pi}
\Pi : S^4_{\mathfrak R_N} \longrightarrow S^4_{\mathfrak R_S}
\end{equation}
by the parametric shifts 
\begin{equation}~\label{shifts}
\theta\mapsto\pi-\theta ~, \quad
\phi\mapsto\phi + \pi~,
\end{equation} 
where thus
\begin{equation}
\Pi (X) = - X~, \quad X\in\mathbb R^5~. 
\end{equation}
Indeed, recalling that $0\leq\theta<\pi/2$ for $\mathfrak R_N$ 
and $\pi/2<\theta\leq\pi$ for $\mathfrak R_S$, it is straightforward 
to verify that~\eqref{shifts} sends every point in $S^4_{\mathfrak R_N}$ 
to an antipodal point at the southern Euclidean wedge $S^4_{\mathfrak R_S}$. 
In particular, the northern observer  is mapped to the southern observer, \emph{viz}. 
$\Pi(\mathcal O_N)= \mathcal O_S$.

The existence of the antipodal map~\eqref{Pi} amounts to effectively describe
the bulk geometry as the 4-sphere 
\begin{equation}\label{S4modPi}
S^4=\big(\mathfrak R_N \cup \mathfrak R_S\big)_E~\big/~ \Pi\,.
\end{equation}

The map~\eqref{Pi} can be restricted to a 3-sphere inside the
4-sphere~\eqref{S4modPi}. In terms of the embedding coordinates
\eqref{embedding}, this is simply done by setting $X^4=0$. 
Alternatively, in terms of the 4-sphere coordinates, this restriction is
obtained by going to the meridian $\phi=0$. In either case, one induces
the symmetry 
\begin{equation}
\Pi \big\vert_{X^4=0} = \pi~,
\end{equation}
where $\pi : S^3\rightarrow S^3$ is the antipodal map 
that acts by the shifts~\eqref{shifts} followed by evaluation at $\phi=0$. 
It is direct to verify that 
\begin{equation}
\pi(X)=-X ~, \quad X\in\mathbb R^4~.
\end{equation} 
We identify the above map with the boundary antipodal 
map~\eqref{anti}.

\paragraph{Entanglement entropy.}
The entanglement entropy between the northern and southern Rindler wedges $\mathfrak R_N$ and $\mathfrak R_S$ can be computed from a purely bulk perspective, without requiring holography. For this, we shall only assume the existence of some quantum gravity theory on dS space whose infrared limit is Einstein gravity. In this limit, the quantum gravity partition function 
can be approximated by the Euclidean on-shell gravity action. 
Importantly, the antipodal map~\eqref{Pi} implies that the bulk 
partition function can be taken to have support on a single 4-sphere.

Following the same logic as in Section~\ref{Sec:4.1}, we take the bulk manifold to be the 4-sphere~\eqref{S4modPi} (which we recall is built up from the two Rindler wedges $\mathfrak R_S$ and $\mathfrak R_N$ modulo the antipodal map~\eqref{Pi}). Next, we choose a Cauchy-like slice whose modular evolution reconstruct the full bulk 4-sphere. We take this to be   
\begin{equation}\label{C4}
\mathcal C_\Phi= B_S^3 \cup B_N^3\,, 
\end{equation}
where $B_S^3$ denotes the 3-ball whose boundary is the 2-sphere $\Sigma_S$. This is obtained by acting with the antipodal map~\eqref{Pi} on $\mathfrak R_S\cong S^4$ and restricting the azimutal angle to a suitable fix value, say $\Phi_0$. That is
\begin{equation}\label{B3}
B^3_S := \big(\mathfrak R_S / \Pi \big) \big\vert_{\Phi_0}~,\quad
\partial B_S^3=\Sigma_S\,,
\end{equation} 
and likewise for $\mathcal D^3_N$. Thus, $\mathcal C_\Phi$ is a 3-sphere constructed by gluing together the 3-balls $B^3_S$ and $B^3_N$ along their boundaries $\Sigma_S$ and $\Sigma_N$, respectively (recall that $\Sigma_S$ and $\Sigma_N$ are the set of fixed points of the bulk $\mathbb Z_q$ action, and they are both 2-spheres in the Euclidean geometry).

Treating $B^3_S$ and $B^3_N$ as two entangled and complementary subsystems, we can now compute the corresponding entanglement entropy. This is given by
\begin{equation}\label{Eq1}
\mathcal S_{\rm E}
= \lim_{q\to1}\frac{\log {\rm Tr}\,\hat\rho_{B^3_S}^{\,q}}{1-q}
= -\partial_q \log {\rm Tr}\,\hat\rho_{B^3_S}^{\,q}\big|_{q=1} ~.
\end{equation}
where the reduced density matrix 
\begin{equation}
\hat\rho_{B^3_S}:={\rm Tr}_{B^3_N}(\hat\rho_{\mathcal C_\Phi})~,
\end{equation}
Here, by means of the replica method, we compute
\begin{equation} 
{\rm Tr}\,\hat\rho_{B^3_S}^{\,q} 
=\frac{\mathcal Z_{\rm QG} [S_q^4]}{(\mathcal Z_{\rm QG} [S^4])^q}~,
\end{equation}
where we have denoted the branched cover of the 4-sphere by $S^4_q$. Taking the logarithm and using the semiclassical approximation $\mathcal Z_{\rm QG}\approx \exp(-I_E)$, with $I_E$ the on-shell gravity action~\eqref{Ionshell2} on the 4-sphere $S^4\cong\mathfrak R$ (where the Rindler wedge can be taken to be the the southern or the northern one, without any loss of generality), it follows that
\begin{align}
\log {\rm Tr}\,\hat\rho_{B^3_S}^{\,q}
=\log \mathcal Z_{\rm QG}[S^4_q]- q\log\mathcal Z_{\rm QG}[S^4]  
\approx -q \Big(I_E[S^4/\mathbb Z_q]- I_E[S^4] \Big)\,.
\end{align}
In the above, we have made use of the locality of the gravity
action to write $I_E[S^4_q]=qI_E[S^4/\mathbb Z_q]$. 
The manifold $S^4/\mathbb Z_q$ is by construction an Eucludian 
Rindler wedge plus the corresponding defect, say $\Sigma$ (which could be either $\Sigma_S$ or $\Sigma_N$). The on-shell value of the action is thus
that of~\eqref{Ionshell2}, with tensionless limit $I_{E}\approx-\pi\ell^2/G_4$. 
Altogether gives 
\begin{align}
\log {\rm Tr}\,\hat\rho_{B^3_S}^{\,q}=2(1-q)\, \frac{A_{\Sigma}}{4G_4}\,,
\end{align} 
and therefore the entanglement entropy~\eqref{Eq1} is given by
\begin{equation}\label{RTdS}
\mathcal S_{\rm E} = \frac{2A_{\Sigma}}{4G_4}=2{\mathcal S}_{dS}\,.
\end{equation}
Observe that the above result differs by a factor of 2 with the Gibbons--Hawking entropy~\eqref{dSS}. We interpret the origin of this extra factor in the extended bulk description of our construction.  Indeed, since the full bulk geometry includes two minimal surfaces, both having the area of a 2-sphere, then $2A_{\Sigma}=:{\rm Area} (\mathcal F)$, where $\mathcal F=\Sigma_S\cup\Sigma_N$ denotes the set of fixed points of the $\mathbb Z_q$ action on $\mathfrak R_S\cup\mathfrak R_N$. Thus, the entropy~\eqref{RTdS} can be recast as the area law
\begin{equation}
\mathcal S_{\rm E} = \frac{{\rm Area} (\mathcal F)}{4G_4} \,.
\end{equation}
This interpretation is compatible with the recent proposal of~\cite{Arias:2019zug}, in which each of the minimal surfaces $\Sigma_S$ and $\Sigma_N$ encode by itself the degrees of freedom (characterized by a central charge proportional to $(1-q^{-1})$ which thus vanishes in the tensionless limit) that give rise to the Gibbons--Hawking entropy; in our current setup, these two minimal surfaces are fully overlapped along the equator of the Cauchy-like slice~\eqref{C4}, and it is thus expectable an enlargement of the total number of microstates localized within the entangling surface, originating the extra factor of 2 in~\eqref{RTdS}. 

\section{Discussions}\label{Sec:5}
In this paper, we have studied the holographic entanglement between the two disconnected conformal boundaries of dS space $\mathcal I^+$ and $\mathcal I^-$, as well as the bulk entanglement between the two causally disconnected Rindler wedges $\mathfrak R_S$ and $\mathfrak R_N$. 

In the former case, we have assumed an holographic duality between quantum gravity in ${\rm dS}_4$ space and two copies of a conformal field theory on the 3-sphere (one copy for each boundary). In the low energy limit, this duality links the CFT partition function with the Euclidean on-shell gravity action as  
$\mathcal Z_{\rm CFT}[S^3]\approx \exp\big(-\tfrac 12 I_{E}[\mathfrak R_S]\big)$ (we recall that this relation has been previously proposed, in a different context, in~\cite{Maldacena:2011mk}). 

In order to compute the holographic entanglement entropy between the two boundaries, we have taken the boundary 3-sphere to by defined as $S^3= (\mathcal I^{-}\cup\mathcal I^{+} )~\big/~\pi$, where $\pi$ is the antipodal map~\eqref{anti}.
We have further chosen a Cauchy-like surface within the boundary $S^3$ given by the 2-sphere that results from gluing together two disks whose boundaries coincide with the $S^1$ set fixed points of the $S^3/\mathbb Z_q$ orbifold, as indicated in Equation~\eqref{Clike}. Consequently and by extending the boundary replica symmetry into the bulk to compute the on shell value of the action $I_{E}[\mathfrak R_S/\mathbb Z_q]$ (displayed in Equation~\eqref{IEtotal}), the entanglement entropy between the two disks---with the entanglement surface located at the equator of the Cauchy-like surface, as depicted in Figure~\hyperlink{Fig:5}5---correctly reproduces the Gibbons--Hawking entropy.

As for the bulk entanglement between the two interior Rindler wedges, we have only assumed the existence of some quantum gravity theory on dS space whose infrared limit is Einstein gravity. In this case, we have treated the bulk manifold as the 4-sphere~$S^4=(\mathfrak R_S\cup\mathfrak R_N)\big/\Pi$, where $\Pi$ is the bulk antipodal map~\eqref{Pi} (and the Rindler wedges are implicitly Euclideanized). Following the same rationale as before, we picked a Cauchy-like surface given in this case by the 3-sphere that results from gluing together two 3-balls whose boundaries coincide with the set of fixed points of the $S^4/\mathbb Z_q$ orbifold; see Equations~\eqref{C4} and~\eqref{B3}. The latter set of fixed points, that we have denoted by $\mathcal F=\Sigma_S\cup\Sigma_N$, are precisely the pair of codimesion two minimal surfaces constructed in Section~\ref{Sec:3}. Calculation of the entanglement entropy between the two 3-balls that bipartition the  Cauchy-like slice (again, with the entanglement surface located at the equator of the $S^3$), yields an entanglement entropy that obeys the area law $S_{\rm E}=\frac14{\rm Area}(\mathcal F)$.

The above results seem to indicate that the notion of entanglement in dS space follows directly from the connectedness properties of the spacetime manifold. In this sense, our findings align with previous ideas~\cite{Kitaev:2006aa, Salton:2017aa, Melnikov:2018aa} in which entanglement follows from topology. Regarding the physical interpretation of our results as well as directions for future research, we shall now elaborate on what we consider are the most relevant ideas underlying our construction and whose generalization and further study might be of interest.
\paragraph{Bulk origin of boundary replica symmetry and 
boundary antipodal map.}
From the bulk perspective there exists two fundamental elements in our construction,  namely a discrete $\mathbb Z_q$ symmetry and an antipodal map. 
We interpret the bulk $\mathbb Z_q$ action as the response of the background geometry to the presence of \emph{massive, non-probe observers}~\cite{Arias:2019zug}; the back-reaction of such observers---that breaks the $SO(3)$ symmetries of the cosmological horizon down to a $U(1)$ symmetry---induces a pair of codimension two minimal surfaces defined by the set fixed points of the ${\rm dS_4}/\mathbb Z_q$ orbifold.  
When restricted to the boundaries, the bulk $\mathbb Z_q$ symmetry translates into a boundary replica symmetry whose fixed points extends into the fixed points of the bulk $\mathbb Z_q$ action as $S^1\hookrightarrow S^2$.

The bulk antipodal map~\eqref{Pi} sends every point in an Euclidean Rindler wedge into a point in the antipodal wedge (in particular, it maps the northern observer $\mathcal O_N$ to the southern observer $\mathcal O_S$) and it amounts to write the bulk gravity partition function with support on a single 4-sphere. 
When restricted to the boundaries, it translates into the boundary antipodal map~\eqref{anti} that sends every point on the past 3-sphere to an antipodal point on the future 3-sphere. 

The existence of both, the bulk $\mathbb Z_q$ symmetry and bulk antipodal map, indicates that the boundary replica symmetry and boundary antipodal map are not truly fundamental as they both arise as restrictions to the boundary of a bigger bulk symmetry and map.

\paragraph{De Sitter energy and holography.}
In four dimensions, pure dS space has zero energy~\cite{Balasubramanian:2002aa}. This can be visualized in a simple way from the canonical thermodynamic relation $\beta F=\beta E -\mathcal S_{dS}$, where $\beta=T^{-1}_{dS}=2\pi\ell$. Here, the Gibbs free energy is defined by the (quantum) gravity partition function and it is well approximated, in the semiclassical limit, by the on-shell gravity action, that is $\beta F=-\log\mathcal Z_{\rm QG}\approx I_{\rm E}$. 
Since the on-shell value of the Euclidean Einstein--Hilbert action--\emph{without boundary terms}--equals minus the value of the dS entropy, $I_{\rm E}\approx -\mathcal S_{dS}$, it follows that $E=0$. However, based on the holographic description of a massive inertial observer proposed in Section~\ref{Sec:4.1}, a non-vanishing notion of energy may be defined.

Indeed, the holographic duality between an inertial observer and a thermofield double state $\lvert \mathcal O_S \rangle\in\mathcal H_{\mathcal I^-}\otimes \mathcal H_{\mathcal I^+}$ implies that the observer density matrix $\rho_{\mathcal O}:= \lvert \mathcal O_S \rangle \langle \mathcal O_S \vert$ is thermal. Furthermore, the entanglement entropy $\mathcal S_{\rm E} = -{\rm tr}(\rho_{\mathcal O} \log \rho_{\mathcal O})$ obeys the canonical relation $\beta F=\beta\langle H\rangle - \mathcal S_{\rm E}$. In this case, the thermal free energy is defined in terms of the boundary field theory partition function $\beta F=-\log\mathcal Z_{\rm CFT}$ and $H$ is the modular Hamiltonian. In the low energy limit, the partition function is given by the Euclidean on-shell gravity action $\beta F\approx  I_{E}[\mathfrak R_S/\mathbb Z_q]$, which now \emph{includes a Nambu--Goto boundary term} (see Equations~\eqref{IEtotal}). Then, using the fact that the entanglement entropy equals dS entropy, one finds 
\begin{equation}
\langle H\rangle 
= \Big(1 - \frac1q\Big) \frac{\ell}{G_4}~.
\end{equation}
This suggest the possibility of defining the four-dimensional dS energy in terms of the expectation value of the boundary modular Hamiltonian $E:=\langle H\rangle$. Note that the tensionless limit $q\to1$ is equivalent to $E=0$ and implies the absence of boundary terms.

\paragraph{AdS space and entanglement}
The non-holographic calculation of the entanglement entropy between the two Rindler wedges in dS space can arguably be extended to AdS space. Essentially, the idea is to glue together two copies of four-dimensional AdS space along their conformal boundary and formally treat each of these copies as a Rindler wedge. This construction can be implemented by extending the radial AdS coordinate as to run over the whole real line (from minus infinity to infinity). The gluing procedure enhances a $\mathbb Z_2$ symmetry and, as a result, the two fully overlapped conformal boundaries become a single domain wall. 
Next, foliating the two AdS copies as $H_2\times_w H_2$ (this is simply the hyperbolic version of the $S^2\times_wS^2$ foliation used in this paper, where $H_2$ denotes the two-dimensional hyperbolic space and $w$ is an hyperbolic warp factor), one locates a pair of causally disconnected massive observers at the fixed points of the orbifold singularities $H_2/\mathbb Z_q$. The two observers are separated by an infinite distance with the conformal boundary located in the middle.
Taking a partial trace over one of the two AdS copies, one obtains the entanglement entropy $\mathcal S_{\rm E}={\rm Area}(\mathcal F)/4G_4$, where ${\rm Area}(\mathcal F)$ is the area of the two minimal surfaces defined by set of fixed points of the two orbifolds $H_2/\mathbb Z_q$ (one orbifold for each AdS copy). We plan to refine and present these ideas in a separate work.

\paragraph{Multiple boundaries and generalized thermofield states.}
The Penrose diagram of dS space (see Figure~\hyperlink{Fig:3}3) is formally the same to that of the eternal  AdS black hole~\cite{ISRAEL1976107,Maldacena:2001kr}, with the two asymptotic AdS boundaries replaced by the two antipodal dS observers, and the black hole singularities replaced by the future and past infinities. 
Motivated in part by this simple observation, we have proposed that an inertial observer in dS space can be holographically modeled as thermofield double state. 

The above dualities relating (maximally extended A)dS gravity and 
thermofield double CFT states are formulated on manifolds with two boundaries. It is thus natural to consider manifolds with an arbitrary number of disconnected and entangled boundaries, say $b$ of them, and further propose that massive 
observers can be holographically described by generalized thermofield states  
\begin{equation}\label{GTS}
\vert \mathcal O\rangle \sim \sum_{n} e^{-\beta E_n/b}
\vert n; q\rangle_1\otimes\vert n; q\rangle_2\otimes\cdots\otimes\vert n; q\rangle_b~. 
\end{equation}
For recent and related work on these ideas see~\cite{Balasubramanian:2014aa, Bao:2018aa, Dwivedi:2018aa, Peach:2017aa}.
The important point is that the observer back-reaction breaks some of the bulk symmetries via the singular $\mathbb Z_q$ action which, upon restriction to the boundaries, reincarnates as a boundary replica symmetry. The orbifold parameter thus enters as a quantum number labeling the fundamental states that build up the composite observer state~\eqref{GTS}, where the probe limit $q\to1$ represents only a single point in the full moduli of the bulk theory.

\section*{{\large Acknowledgements}}
It is a pleasure to thank G.~Anastasiou, R.~Aros, F.~Rojas, G.~Valdivia, B.~Vallilo and J.~Zanelli for interesting discussions related to the topic of this paper and K.~Narayan and M.-I. Park for correspondence and references. 
We are especially indebted to I.~Araya, G.~Arenas-Henriquez, R.~Olea and D.~Rivera-Betancour for reading our work, detailed discussions clarifying aspects of entanglement entropy, and many suggestions for improving the presentation.

{\sc Ca}~and~{\sc Fd} are Universidad Andres Bello 
({\sc Unab) PhD} and {\sc MSc} Scholarship holders, 
respectively.  Their work is supported by the Direcci\'on General de 
Investigaci\'on ({\sc Dgi-Unab}). 
The work of {\sc Ps} is partially supported by {\sc Fondecyt} Regular
grants N$^{\rm o}$1140296 and N$^{\rm o}$1151107, and 
{\sc Conicyt} grant {\sc Dpi} 2014-0115.

%
\bibliographystyle{JHEP}
\bibliography{dSeeREF}
\end{document}